\documentstyle[preprint,aps]{revtex}
\draft
\begin{document}
\newcommand{\be}{\begin{equation}}
\newcommand{\ee}{\end{equation}}
\newcommand{\bea}{\begin{eqnarray}}
\newcommand{\eea}{\end{eqnarray}} 

\title{Periodic Orbits in Polygonal  
Billiards \thanks{
{\it Nonlinearity and Chaos in the Physical Sciences}, Pramana
(Special Issue)}}

\author{Debabrata Biswas \cite{email,permanent}}

\address{Center for Chaos and Turbulence Studies,
Niels Bohr Institute \\
Blegdamsvej 17, 
Copenhagen $\O$, Denmark}
\maketitle
\vskip 0.2 in
\begin{abstract}
\par
We review some properties of periodic orbit families in
polygonal billiards and discuss in particular a sum
rule that they obey. In addition, we provide algorithms to determine periodic
orbit families and present numerical results that shed
new light on the proliferation law and
its variation with the genus of the invariant surface. 
Finally, we deal with correlations in the length spectrum 
and find that long orbits display Poisson fluctuations.  

\end{abstract}

\vskip 0.75 in

\section{Introduction}

\par Billiards are an interesting and well studied class of 
hamiltonian systems that display a wide variety of dynamical
behaviour depending on the shape of the boundary. 
The ray equations commonly considered follow from a short wavelength
expansion of the Schr\"{o}dinger equation with Dirichlet (or Neumann)
boundary conditions. A particle thus moves freely between collisions
at the boundary where it suffers specular reflection.
As an example, a square
billiard generates regular dynamics that is restricted to a
torus in phase space due to the existence of two well behaved
constants of motion. In contrast, generic trajectories in the
enclosure formed by three intersecting discs 
explore the entire constant energy
surface and hence the system is ergodic. Moreover, orbits
that are nearby initially move exponentially apart with time
and hence the system is said to be metrically chaotic.

\par Polygonal billiards are a class of systems
whose properties are not as well known. The ones accessible
to computations are rational angled polygons and they generically 
belong to
the class of systems referred to as pseudo-integrable \cite{PJ}.
They possess two constants of motion like their integrable
counterparts \cite{rem0} but
the invariant surface $\Gamma$ is topologically equivalent to a sphere
with multiple holes \cite{PJ} and not a torus. Such systems
are characterised by the genus $g$ where $2g$ is the
numbers of cuts required in $\Gamma$ to produce a singly 
connected region (alternately, $g$ is the number of holes in
$\Gamma$). As an example, consider the 1-step billiard 
in Fig.~1. For any trajectory,  
$p_x^2$ and $p_y^2$ are conserved. The
invariant surface consists of four sheets (copies) corresponding to the 
four possible momenta, ($\pm p_x, \pm p_y$) that it can have 
and the edges of these
sheets can be identified such that the resulting surface has the
topology of a double torus.

\par The classical dynamics no longer has the simplicity of an
integrable system where a transformation to {\it action} and {\it angle}
co-ordinates enables one to solve the global evolution equations 
on a torus. On the other hand, the
dynamics is non-chaotic with the only interesting feature occurring
at the singular vertex with internal angle $3\pi/2$. Here,
families of parallel rays split and 
traverse different paths, a fact that limits the extent
of periodic orbit families. This is in contrast to integrable 
billiards (and to the $\pi/2$ internal angles in Fig. 1)
where families of rays do not see the vertex and continue
smoothly.  

\par We shall focus here on the periodic orbits of
such systems for they form the skeleton on which
generic classical  motion is built. They are also the
central objects of modern 
semiclassical theories \cite{MC} which provide a 
duality between the quantum spectrum and the 
classical length and stabilities of periodic
orbits. In polygonal billiards, primitive periodic orbits 
can typically be classified under two categories 
depending on whether they suffer even or odd 
number of bounces at the boundary. In both cases
however, the linearised flow is marginally unstable
since the Jacobian matrix, $J_p$, connecting the transverse
components ($u_{\perp}(t + T_p) = J_p\;u_{\perp}(t)$ where $
u_\perp = (q_\perp,p_\perp)^T$ and $T_p$ is the time period
of the orbit)
has unit eigenvalues. However, when the number of
reflections, $n_p$, at the boundary is even, the orbit
occurs in a 1-parameter family while it is isolated
when $n_p$ is odd. This follows from the fact 
that the left (right) neighbourhood of an  
orbit becomes the right (left) neighbourhood
on reflection so that an initial neighbourhood
can never overlap with itself after odd number
of reflections \cite{PJ}. 

\par While isolated orbits are important and need
to be incorporated in any complete  {\it periodic
orbit theory}, a family of identical orbits has 
greater weight and such families proliferate faster than
isolated orbits \cite{gutkin}. Besides, in a number of cases
including the L-shaped billiard of Fig.~1, isolated
periodic orbits are altogether absent since the
initial and final momenta coincide only if the
orbit undergoes even number of reflections.
For this reason, we shall consider families of
periodic orbits in this article (see Fig.~(2) for
an example). 

\par Not much is however known
about the manner in which they are organised and the 
few mathematical results that exist \cite{gutkin}
concern the asymptotic properties of their proliferation
rate. For a sub-class of rational polygons where the
vertices and edges lie on an integrable lattice (the
so called almost-integrable systems \cite{gutkin}), these
asymptotic results are exact. It is known for example 
that the number of periodic orbit families, $N(l)$, 
increases quadratically with length, $l$, as $l \rightarrow 
\infty$. For general rational polygons, rigorous results
show that $c_1 l^2 \leq N(l) \leq c_2 l^2$ for sufficiently
large values of $l$ \cite{gutkin,masur} while in case of  
a regular polygon $P_n$ with $n$ sides, it is known that
$N(l) \simeq c_n l^2/A$ \cite{veech} where $c_n$ is
a number theoretic constant and $A$ denotes the area
of $P_n$.
Very little is known however about other aspects such as the
sum rules obeyed by periodic orbit families in 
contrast to the limiting cases of integrable and
chaotic behaviour where these have been well studied.
Besides, it is desirable to learn about the variation
of the proliferation rate as a function of the genus
for this should tell us about the transition to 
chaos in polygonal approximations of chaotic 
billiards \cite{vega,cc}.

\par We shall concern ourselves primarily with
a basic sum rule obeyed by periodic orbits arising 
from the conservation of probability. This leads
to the proliferation law, $N(l) = \pi b_0 l^2/ \langle a(l)
\rangle$ where $b_0$ is a constant. The quantity 
$\langle a(l) \rangle$ is the average area occupied 
by all families of periodic orbits with length less than
$l$ and is not a constant unlike integrable billiards.
We provide here the some numerical results on how
$\langle a(l) \rangle$ changes with the length of the
orbits and the genus of the invariant surface. While
this does not allow us to make quantitative predictions,
the qualitative behaviour sheds new light on the
proliferation law for short orbits and its variation 
with the genus, $g$ of the invariant surface.

\par Finally, we shall also study correlations
in the length spectrum of periodic orbits. The numerical
results provided here are the first of its kind
for generic pseudo-integrable billiards and 
corroborate theoretical predictions provided
earlier \cite{PLA}.

\par The organisation of this paper is as follows.
In section \ref{sec:sum-rule}, we provide the 
basic sum rule and the proliferation law in 
pseudo-integrable systems. In section \ref{sec:num-po},
we discuss algorithms to determine periodic orbit
families in generic situations. This is 
followed by a numerical demonstration of the
results obtained in section \ref{sec:sum-rule}
as well an exploration of the $area \;law$.
Correlations are discussed in sections \ref{sec:corrl}
together with numerical results.
Our main conclusions are summarised in section \ref{sec:concl}.

\section{The Basic Sum Rule}
\label{sec:sum-rule}

\par The manner in which periodic orbits organise themselves
in closed systems is strongly linked to the existence of
sum rules arising from conservation laws. For example, the
fact that a particle never escapes implies that for
chaotic systems \cite{PCBE,HO}

\be \langle \sum_p \sum_{r=1}^{\infty} {T_p \delta(t-rT_p)\over
\left |{ \det({\bf 1} - {{\bf J}_p}^r)} \right |} \rangle = 1\label{eq:hbolic}
\ee

\noindent                                
where the summation over p refers to all primitive periodic
orbits, $T_p$ is time period, ${\bf J}_p$ is the stability
matrix evaluated on the orbit and the symbol $\langle . \rangle$
denotes the average value of the expression on the
left. Since the periodic orbits are
unstable and isolated, $\left | \det({\bf 1} - {{\bf J}_p}^r) \right |
\simeq e^{\lambda_p rT_p}$, where $\lambda_p$ is the Lyapunov
exponent of the orbit. The exponential proliferation of orbits
is thus implicit in eq.~(\ref{eq:hbolic}).

\par A transparent derivation of Eq.~(\ref{eq:hbolic}) follows 
from the classical evolution operator  

\be
L^t{\circ}\phi({\bf x}) = \int {\bf dy}\; \delta({\bf x} -
{\bf f}^t({\bf y}))\;\phi({\bf y}) \label{eq:def1}
\ee

\noindent
where $L^t$ governs the evolution of densities $\phi({\bf x})$,
{\bf x} = ({\bf q,p}) and ${\bf f}^t$ refers to the flow in
the full phase space.
We denote by $\Lambda_n(t)$ the eigenvalue
corresponding to an eigenfunction $\phi_n({\bf x})$ such that
$L^t{\circ}\phi_n({\bf x}) = \Lambda_n(t) \phi_n({\bf x})$.
The semi-group property, $L^{t_1}\circ L^{t_2}
= L^{t_1 + t_2}$, for continuous time implies that
the eigenvalues $\{\Lambda_n(t)\}$ are of the form $\{e^{\lambda_n t}\}$.
Further, for hamiltonian flows, Eq.~(\ref{eq:def1}) implies that
there exists a unit eigenvalue corresponding
to a uniform density so that $\lambda_0 = 0$. For
strongly hyperbolic systems, $\lambda_n = -\alpha_n + i\beta_n, n > 1$
with a negative real part implying  that

\be
{\rm Tr}\; L^t = 1 + \sum_n \exp\{-\alpha_nt + i\beta_nt\} \label{eq:trace0} 
\ee

\noindent
Eq.~(\ref{eq:hbolic}) is thus a restatement of Eq.~(\ref{eq:trace0})
with the trace expressed in terms of periodic orbit stabilities
and time periods.

\par For polygonal billiards, appropriate modifications
are necessary to take account of the fact that the flow
is restricted to an invariant surface that is two
dimensional. Further, since  classical considerations do 
not always yield the spectrum $\{\lambda_n(t)\}$, we shall
take resort to the semiclassical trace formula which
involves periodic orbit sums similar to the kind that
we shall encounter in the classical case.

\par Before considering the more general case of pseudo-integrable
billiards, we first introduce the appropriate {\it classical}
evolution operator for integrable systems. This is easily
defined as  

\be
L^t{\circ}\phi(\theta_1,\theta_2) =
\int d\theta_{1}'d\theta_{2}'\;
\delta (\theta_1 - \theta_{1}'^t)\delta (\theta_2 - \theta_{2}'^t)
\;\phi(\theta_1',\theta_2') \label{eq:prop}
\ee

\noindent
where $\theta_1$ and $\theta_2$ are the angular coordinates on the
torus and evolve in time as $\theta_i^t = \omega_i (I_1,I_2)t +
\theta_i$ with $\omega_i = \partial H(I_1,I_2)/\partial I_i $ and
$I_i = {1\over 2\pi}\oint_ {\Gamma_i} {\bf p.dq} $.
Here $\Gamma _i, i = 1,2$ refer to the
two irreducible circuits on the torus and ${\bf p}$ is the momentum
conjugate to the coordinate ${\bf q}$.
               
\par  It is easy to see that the eigenfunctions
$\{ \phi_n(\theta_1,\theta_2)\}$ are such that
$\phi_n(\theta_1^t,\theta_2^t) = \Lambda_n(t) \phi_n(\theta_1,\theta_2)$
where $\Lambda_n(t) = e^{i\alpha_n t}$. On demanding that
$\phi_n(\theta_1,\theta_2)$
be a single valued function of $(\theta_1,\theta_2)$, it follows
that $\phi_{\bf n}(\theta_1,\theta_2) = e^{i(n_1\theta_1 + n_2\theta_2)}$
where ${\bf n} = (n_1,n_2)$  is a point on the integer lattice.
Thus the eigenvalue,
$\Lambda_{\bf n}(t) = {\rm exp}\{it(n_1\omega_1 + n_2\omega_2)\}$.

\par To illustrate this, we consider a rectangular billiard
where the
hamiltonian expressed in terms of the actions,
${I_1,I_2}$ is $H(I_1,I_2) = \pi^2(I_1^2/L_1^2 + I_2^2/L_2^2)$
where $L_1,L_2$ are the lengths of the two sides.
With $I_1 = \sqrt{E}L_1\cos(\varphi )/\pi$ and
$I_2 = \sqrt{E} L_2\sin(\varphi )/\pi$, it is easy to
see that at a given energy, $E$, each torus is parametrised by a
particular value of $\varphi$. Thus 

\be
\Lambda_{\bf n}(t) =
e^{i2\pi t\sqrt{E}(n_1\cos(\varphi)/L_1 + n_2\sin(\varphi)/L_2)} 
\label{eq:rect}
\ee

\noindent
and the spectrum is continuous. The trace thus involves an
integration over $\varphi$ as well as a sum
over all $n_1,n_2$ :

\be
{\rm Tr}~L^t = \sum_{\bf n} \int_{-\pi - \mu_n}^{\pi - \mu_n} d\varphi\;
e^{il\sqrt{E_{\bf n}}\sin(\varphi + \mu_{\bf n})}
= 2\pi \sum_{\bf n} J_0(\sqrt{E_{\bf n}}l)  \label{eq:bessel}
\ee

\noindent
where $J_0$ is a Bessel function,  
$l = 2t\sqrt{E}$, $\tan(\mu_{\bf n}) = n_1L_2/(n_2L_1)$
and $E_n = \pi^2(n_1^2/L_1^2 + n_2^2/L_2^2)$. 
On separating out ${\bf n} = (0,0)$ from the rest and restricting
the summation to the first quadrant of
the integer lattice, it follows that 

\be
{\rm Tr}~L^t = 2\pi  + 2\pi N \sum_n J_0(\sqrt{E_n}l)
\label{eq:rect3}.
\ee  

\noindent
where $N=4$. Note that the first term on the right merely
states the fact that there exists a unit eigenvalue 
on every torus labelled by $\varphi$. Also, though we
have not invoked semiclassics at any stage, the spectrum
$\{E_n\}$ corresponds to the Neumann spectrum of the
billiard considered. We shall subsequently show that
this is true in general and for now it remains to
express the trace of the evolution operator in 
terms of periodic orbits.

\par The trace of $L^t$ in the integrable case can
be expressed as 

\be
{\rm Tr}\; L^t = \int d\varphi \int d\theta_1 d\theta_2 \;
\delta (\theta_1 - \theta_{1}^t)\delta (\theta_2 - \theta_{2}^t)
\ee

\noindent
It follows immediately that the only orbits that contribute
are the ones that are periodic. For a rectangular billiard, 
the integrals can be evaluated quite easily and yields 

\be
{\rm Tr}\; L^t = 4 \sum_{N_1} \sum_{N_2} {4L_1L_2\over l_{N_1,N_2}}
\delta(l- l_{N_1,N_2}) \label{eq:rect4}
\ee

\noindent
where $\{N_1,N_2\}$ are the winding numbers on the torus and
label periodic orbits of length $l_{N_1,N_2}$. Using 
Eqns.~(\ref{eq:rect3}) and (\ref{eq:rect4}), it follows that
 
\be
4 \sum_{N_1} \sum_{N_2} {4L_1L_2\over l_{N_1,N_2}}
\delta(l- l_{N_1,N_2}) = 2\pi  + 2\pi N \sum_n J_0(\sqrt{E_n}l)
\label{eq:rect5}
\ee

\noindent
Thus, the dominant non-oscillatory contribution to the
trace that survives
averaging is $2\pi$ and this gives rise to the analogue
of Eq.~(\ref{eq:hbolic})

\be
4 \langle \sum_{N_1} \sum_{N_2} {4L_1L_2\over l_{N_1,N_2}}
\delta(l- l_{N_1,N_2}) \rangle = 2\pi 
\ee

\noindent
The proliferation rate for the rectangular billiard thus
follows from these considerations. 
 
\par These ideas can be generalised for polygonal billiards
that are pseudo-integrable even though the structure of
the invariant surface no longer allows one to use
{\it action} and {\it angle} variables.

\par For both integrable and pseudo-integrable polygonal
billiards,
the dynamics in phase space can be viewed in a singly connected region by
executing $2g$ cuts in the invariant surface and identifying
edges appropriately. At a given energy, the motion is
parametrised by the angle, $\varphi$, that a trajectory
makes with respect to one of the edges. As a trivial example,
consider the rectangular billiard.
The singly connected region is a larger rectangle consisting
of four copies corresponding
to the four directions that a trajectory can have and these
can be glued appropriately to form a torus \cite{keller}.
As a non-trivial
example, consider the L-shaped billiard of Fig.~(1) which is
pseudo-integrable with its invariant surface having, $g=2$.
Alternately, the surface can be represented by a singly connected
region in the plane (see Fig.~(3)) and consists of four copies
corresponding to the four possible directions an orbit can
have and these are glued appropriately.
A trajectory in phase space thus consists of parallel segments at an
angle $\varphi$ measured for example with respect to one of the sides.
It will be useful to note at this point that the same trajectory
can also be represented by parallel segments at angles
$\pi - \varphi$, $\pi + \varphi$ and $2\pi - \varphi$.
In general, the number of
directions for representing a trajectory equals the number of copies, $N$,
that constitute the invariant surface.

\par The classical propagator on an invariant surface parametrised
by $\varphi$ is thus

\be
L^t (\varphi) {\circ}\phi({\bf q}) = \int d{\bf q'}\;
\delta({\bf q} - {\bf q}'^t(\varphi))
\;\phi({\bf q}') \label{eq:prop2}
\ee

\noindent
where ${\bf q}$ refers to the position in the singly connected region
and ${\bf q}'^t(\varphi)$ is the
time evolution parametrised by $\varphi$ as described above.

\par The trace of $L^t (\varphi)$ takes into account all possible
invariant surfaces that exist and hence involves an additional integration
over $\varphi$. Thus

\be
{\rm Tr}~L^t = \int d\varphi \int d{\bf q} \;\delta({\bf q} -
{\bf q}^t(\varphi)) \label{eq:trace3}
\ee

\noindent
Clearly the only orbits that contribute are the ones that are
periodic. Further, the {\bf q} integrations are simpler if we
transform to a local coordinate system with one component
parallel to the trajectory and the other perpendicular.
Thus $\delta_{\|}(q_\| - q_\| ^t) = {1\over v}\delta (t-rT_p) $
where $v$ is the velocity, $T_p$ is the period of the
orbit and $r$ is the repetition number. Similarly, for
an orbit of period $T_p$ parametrised by the angle
$\varphi_p$, $\delta_{\bot} (q_\bot - q_\bot ^t) =
\delta (\varphi - \varphi_p)/{\left |\partial q_\bot/\partial
\varphi \right |_{\varphi = \varphi_p} }$ where
$\left |\partial q_\bot/\partial \varphi
\right |_{\varphi = \varphi_p} = rl_p$ for marginally
unstable billiards. Putting these results together and
noting that each periodic orbit occurs in general at
$N$ different values of $\varphi$,  we finally have

\be
{\rm Tr}~L^t = \sum_n \Lambda_n(t) = N~\sum_p \sum_{r=1}^{\infty}
 {a_p\over rl_p}\delta(l-rl_p) \label{eq:trace4}
\ee
                                                                    
\noindent
where $l = tv$ and the summation over $p$ refers to all primitive
periodic orbit families with length $l_p$ and  occupying an area $a_p$.
Note that Eq.~(\ref{eq:rect4}) is a special case of 
Eq.~(\ref{eq:trace4}) which holds for both integrable
and non-integrable polygonal billiards.

\par It is possible to re-express the periodic
orbit sum in Eq.~(\ref{eq:trace4}) starting with the appropriate
quantum trace formula \cite{neglect-isolated1}

\be
\sum_n \delta (E - E_n) = d_{av}(E) + {1\over \sqrt{8\pi^3}}
\sum_p \sum_{r=1}^{\infty} {a_p\over \sqrt{krl_p}}\cos(krl_p -
 {\pi\over 4} - \pi rn_p) \label{eq:richens}
\ee

\noindent
Here $d_{av}(E)$ refers to the average
density of quantal eigenstates, $k = \sqrt{E}$,
$l_p$ is the length of a primitive
periodic orbit family. The phase $\pi r n_p$ is set to zero 
while considering the Neumann spectrum while in the
Dirichlet case, $n_p$ equals the number of bounces
that the primitive orbit suffers at the boundary.
For convenience,
we have chosen $\hbar = 1$ and the mass $m=1/2$.
Starting with the function

\be
g(l) =  \sum_n f(\sqrt{E_n}l) e^{-\beta E_n} =
\int_\epsilon^\infty dE\; f(\sqrt{E}l) e^{-\beta E} \sum_n \delta(E-E_n)
\ee

\noindent
where $f(x) = \sqrt{{2\over \pi x}} \cos(x - \pi/4)$
and $ 0 < \epsilon < E_0 $,
it is possible to show using Eq.~(\ref{eq:richens}) that
for polygonal billiards \cite{prl1}

\be
\sum_p \sum_{r=1}^{\infty} {a_p   \over rl_p} \delta (l-rl_p)
 = 2\pi b_0 + 2\pi \sum_n f(\sqrt{E_n}l) \label{eq:myown1}
\ee

\noindent
for $\beta \rightarrow 0^+$. In the above,

\be
b_0 = \sum_p \sum_r {a_p (-1)^{rn_p} \over 4\pi}\int_0^{\epsilon}\; dE
f(\sqrt{E}l)f(\sqrt{E}rl_p)
\ee

\noindent
and is a
constant \cite{prl1,rapid?}.
It follows from Eqns.~(\ref{eq:myown1}) and~(\ref{eq:trace4}) that

\be
{\rm Tr}~L^t = \sum_p \sum_{r=1}^{\infty} { N a_p \over rl_p} \delta (l-rl_p) 
 = 2\pi Nb_0 + 2\pi N \sum_n f(\sqrt{E_n}l)
\label{eq:trace5}.
\ee

\noindent
where $\{E_n\}$ are the Neumann eigenvalues of the system. As in case of 
the rectangular billiard, the oscillatory contributions wash
out on averaging so that 

\be 
\langle \sum_p \sum_{r=1}^{\infty} {a_p \over rl_p} \delta (l-rl_p)
\rangle = 2\pi b_0  \label{eq:basic}
\ee

\noindent
This is the central result of this section and forms the
basic sum rule obeyed by periodic orbit families.

\par A few remarks about this derivation and the magnitude
of $b_0$ are however in order. Eq.~(\ref{eq:trace4}) is
exact for the L-shaped billiard and all other boundaries
which preclude the existence of isolated periodic orbits.
However, even for these shapes, the semiclassical 
trace formula is only an approximation to the exact
density whenever the billiard in question is pseudo-integrable.
In this sense, the sum rule in Eq.~(\ref{eq:basic}) is
not expected to be exact. However, we believe the existence
of higher order corrections (such as diffraction) affects
the magnitude of the constant $b_0$ while preserving
the constant nature of the periodic orbit sum on the
left.

\par In the integrable case it is easy to show from 
other considerations that $b_0 = 1/N$ where $N$ is the 
number of sheets that constitute the invariant surface 
\cite{prl1}. This also follows from Eqns.~(\ref{eq:rect5})
and (\ref{eq:trace5})
since $b_0N = 1$. For pseudo-integrable billiards,
each invariant surface parametrised by $\varphi$
has an eigenvalue, $\Lambda_0(\varphi) = 1$.
Thus the non-oscillatory 
part of the trace should equal $\int \Lambda_0(\varphi)
\; d\varphi$ and to a first approximation
this yields $2\pi$ implying that $b_0 = 1/N$.  
However each singular vertex connects
two distinct points at any angle $\varphi$ and hence 
the integration over $\varphi$ is non-trivial. 
We can therefore state that $b_0$ is approximately
$1/N$ in the pseudo-integrable case while this
is exact in the integrable case. We shall show 
numerically that the magnitude of deviations (from $1/N$)
in the
pseudo-integrable case depends on the existence
of periodic orbit pairs at the singular vertex.
First however, we briefly describe the algorithms
used to determine periodic orbit families.

\section{Algorithms for Determining periodic orbits}
\label{sec:num-po}  

\par Periodic orbits in polygonal billiards are
generally hard to classify. Unlike integrable billiards,
they cannot be described by two integers 
which count the number of windings around the
two irreducible circuits on the torus though
in exceptional cases this can indeed be done.
However, since the invariant surface has
a well-defined genus, it is expected that
a set of integers ${\bf N} = \{N_1,N_2,\ldots,N_{2g}\}$
obeying the relationship

\be
\omega_i = {2\pi N_i \over T_{\bf N}}
\ee

\noindent
can be used to label periodic orbits. Here $\omega_i$
refers to the frequency corresponding to each
irreducible circuit $\Gamma_i$ and depends
on the energy, E and the angle $\varphi$ that
labels each invariant surface.
Note however
that not all points on this multi-dimensional
integer lattice are allowed since there are
constraints and this method of labelling orbits
becomes cumbersome for surfaces of higher
genus. Nevertheless, we illustrate the idea
here for the L-shaped billiard of Fig.~(1).

\par Let the length of the two bouncing
ball orbits in the X-direction be $L_1$ and
$L_2$ respectively and their lengths
in the Y-direction be $L_3$ and
$L_4$. These define the irreducible circuits
for the L-shaped billiard. Thus :

\bea
{v\cos(\varphi) \over 2L_1}& = & {2\pi N_1 \over T_{\bf N}} \nonumber \\
{v\cos(\varphi) \over 2L_2}& = & {2\pi N_2\over T_{\bf N}} \nonumber \\
{v\sin(\varphi) \over 2L_3}& = & {2\pi N_3\over T_{\bf N}} \nonumber \\
{v\sin(\varphi) \over 2L_4}& = & {2\pi N_4\over T_{\bf N}} \label{eq:irc}
\eea

\noindent
This implies that the angle $\varphi$ at which a periodic
orbit can exist is such that 

\be
\tan(\varphi) = {N_3L_3 + N_4L_4 \over N_1L_1 + N_2L_2}
\label{eq:po-angle}
\ee

\noindent
Eqns.~(\ref{eq:po-angle}) and (\ref{eq:irc}) merely express the fact that
any periodic orbit should have integer number of windings 
around the irreducible circuits. Thus, the total displacement along the
X-direction should be $2(N_1L_1 + N_2L_2)$ while
the total displacement in the Y-direction should
be $2(N_3L_3 + N_4L_4)$ where $N_i$ are integers. 
As mentioned before however, not all realizations
of $\{N_i\}$ correspond to real periodic
orbits and the final step consists in checking numerically 
whether a periodic orbit at the angle $\varphi$ (given
by Eq.~(\ref{eq:po-angle})) exists. Note that one
member of each family necessarily resides at one of the
singular vertices and hence it is sufficient to 
verify the existence of this orbit.

\par This method works equally well for other
billiards with steps (see Fig.~(1)) and the
number of integers necessary to describe
orbits increases with the number of steps.
An alternate method which exploits the 
fact that periodic orbits occur in families
is often useful when the irreducible 
circuits are not obvious and this is described
below.

\par Note that a non-periodic orbit originating
from the same point ${\bf q}$ (e.g. the singular vertex)
as a periodic orbit but with a
momentum slightly different
(from the periodic orbit) suffers a net transverse deviation
$q_\perp = (-1)^{n_\varphi}\sin(\varphi - \varphi _p)l_{\varphi}$. Here
$l_{\varphi}$ is the distance traversed by a non-periodic
orbit at an angle $\varphi$ after $n_\varphi$ reflections from the
boundary and $\varphi_p$ is the angle at
which a periodic orbit exists. This provides a correction
to the initial angle and a few iterations are normally
sufficient to converge on a periodic orbit with
good accuracy. In order to obtain all periodic orbits, 
it is necessary to shoot trajectories from every 
singular vertex since one member of each family 
resides at one of these vertices. 

\par  Apart from the length of a periodic orbit,
it is also important to compute the area occupied by the
family. This can be achieved by shooting a single 
periodic trajectory which resides at a singular vertex
and by noting that this orbit lies on the edge
of a family. Thus if the rest of the family 
lies on the left neighbourhood initially, it 
is necessary to determine the perpendicular distance
from a singular vertex to this trajectory every time
the initial neighbourhood lies towards this singular
vertex. The shortest of these perpendicular distances
gives the transverse extent of the family and the
area can thus be computed.

\par These algorithms have been used to generate
the lengths $\{l_p\}$ and areas $\{a_p\}$ of
primitive orbits in the L-shaped billiard as
well as the two and three step billiards. We
present our numerical results in the following
sections.

\section{Numerical results : the sum rule and area law}
\label{sec:num1}

\par We present here our numerical results on the 1-step 
(L-shaped), 2-step and 3-step billiards. Their 
invariant surfaces have genus 2, 3 and 4 respectively
and the quantities we shall  study are the sum rule
derived in section \ref{sec:sum-rule} and the variation
of $\langle a(l) \rangle$ with $l$ and the genus, $g$
of the invariant surface.

\par The sum rule we wish to study can be re-expressed
as  

\be
S(l) = {1\over 2\pi} {\sum_p \sum_r}_{rl_p \leq l} {a_p\over rl_p}
\; \simeq b_0 l   \label{eq:sum0}
\ee

\noindent
and this is plotted in Fig.~(4) for four different 
1-step (L-shaped) billiards. Notice first that in 
each case the behaviour is linear as predicted
by Eq.~(\ref{eq:sum0}). Besides, in 
three of the four cases, the slopes are quite
close ($b_0 \simeq 0.27$) and these correspond to non-degenerate
L-shaped billiards with sides that are unequal
and irrationally related. The one with a substantially
larger slope ($b_0 \simeq 0.32$) is a degenerate
case of a square with a quarter removed and for this 
system there exist a  
substantial number of periodic orbit pairs 
at the same angle on the two  adjacent edges at the 
singular vertex. The differences between the degenerate
and non-degenerate case also persists in other quantities 
as we shall shortly see.

\par We next plot $S(l)$ for non-degenerate examples
of a 1,2 and 3-step billiards in Fig.~(5). The number of orbits
considered are far less due to increased computational
effort though the linear behaviour is obvious in
all three cases. The slopes are again 
close to 0.25 and vary from $b_0 \simeq 0.24$
to $b_0 \simeq 0.28$.

\par Thus, periodic orbits obey a basic sum rule
given by Eq.~(\ref{eq:sum0}) where $b_0$ is a 
constant. Further, the magnitude of $b_0$ is close
to $0.25$ in all cases and the deviations from this
value are larger in the degenerate case where 
periodic orbit pairs exist at the singular vertex.
 
\par The sum rule yields the proliferation
rate of periodic orbits 

\be
N(l) = {\pi b_0 l^2 \over \langle a(l) \rangle }
\label{eq:nl}
\ee

\noindent
where $\langle a(l) \rangle$ is the average area
occupied by periodic orbit families with length
less than $l$. In case of integrable polygons
$\langle a(l) \rangle$ is a constant and equals
$AN$ where $A$ is the area of the billiard while
$b_0 = 1/N$. For pseudo-integrable cases, 
the areas $a_p$ occupied by individual periodic
orbit families generically occupy a wide
spectrum bounded above by $NA$. As  the length
of a family increases, encounters with
singular vertices are more frequent so that
the transverse extent of family decreases.
Quantitative predictions about the behaviour
of $\langle a(l) \rangle$ are however difficult and
we provide here some numerical results.

\par Fig.~(6) shows a plot of $\langle a(l) \rangle / 
NA $ for three different L-shaped billiards one of 
which is the degenerate case presented in Fig.~(4).
The average area increases initially before saturating
for large $l$ in all cases. The normalised saturation
value seems to be independent of the ratio of the
sides as long as they are irrationally related
(we have observed this for other cases not presented
here) but is very different for the degenerate 
example of a square with a quarter removed.

\par A comparison of non-degenerate 1,2 and 3-step
billiards is shown in Fig.~(7). Recall that
the invariant surfaces for these have genus
respectively equal to $2,3$ and $4$. The
saturation observed for the 1-step (L-shaped)
case seems to be a common feature and interestingly
the normalised average saturation value  
decreases. This is however expected since
the number of singular vertices increases
with the genus thereby reducing the average
transverse extent of orbit families at any
given length. 

\par These observations allow us to conclude
that for short lengths, the proliferation law
is sub-quadratic. Asymptotically however,
$N(l) \sim l^2$ as in integrable billiards.
Further, the asymptotic proliferation rate
for billiards with the same area $A$
increases with the genus due to a decrease
in the asymptotic normalised saturation
value of $\langle a(l) \rangle$.

\par These numerical results provide a qualitative
picture of the area law and show that periodic
orbits in polygonal billiards are organised
such that they obey a sum rule. Quantitative
predictions would require a more extensive
numerical study such that empirical laws 
for the saturation value and its variation
with the genus can be arrived at. 

\section{correlations in the Length spectrum}
\label{sec:corrl}

\par Our numerical explorations so far have been
focussed on the average properties of the length
spectrum. We shall now attempt to understand the
nature of the fluctuations and characterise 
their statistical properties. 

\par Fluctuations in the length spectrum are generally
difficult to study from purely classical considerations.
There are notable exceptions however. Fluctuations
in the integrable case can be studied using the 
Poisson summation formula since the lengths 
of orbits are expressed in terms of integers
$\{N_1,N_2\}$ \cite{pramana}. The other extreme is the motion on surfaces of
constant negative curvature where the Selberg trace
formula provides an {\it exact} dual relationship between the      
classical lengths and the eigenvalues of the Laplace-Beltrami
operator \cite{BV}. For other systems, a possible way of studying
fluctuations in the length spectrum lies in inverting the
semiclassical quantum trace formula. For pseudo-integrable billiards,
this has been achieved in section \ref{sec:sum-rule} and
the integrable density of lengths, $N(l)$ can be expressed
as 

\be
N(l) = {\pi b_0 l^2 \over \langle a(l) \rangle } +
{2\pi \over \langle a(l) \rangle} \sum_n \int_0^l dl' \;l'f(\sqrt{E_n}l')
\label{eq:nl-tot}
\ee

\noindent
Statistical properties of the fluctuations can thus be
studied using techniques introduced for the fluctuations 
in the quantum energy spectrum \cite{berry85,PLA}. 

\par The correlations commonly studied are the nearest neighbour
spacings distribution, $P(s)$ and a two-point correlation referred to
as the spectral rigidity, $\Delta_3(L)$. The rigidity measures
the average mean square deviation of the staircase function
$N(l)$ from the best fitting straight line over $L$ 
mean level spacings. For a normalised 
(unit mean spacing) Poisson
spectrum, $P(s) = e^{-s}$ while $\Delta_3(L) = L/15$. These
features are commonly observed in the quantum energy spectrum
of integrable systems as well as in their length spectrum.
For chaotic billiards, the quantum energy spectrum generically
produces non-Poisson statistics while the length spectrum
correlations are Poisson for long orbits at least over short
ranges \cite{prl2}.
There are however deviations that can be observed in
$\Delta_3(L)$ for short orbits or over longer ranges in the
spectrum \cite{prl2}. With this background, we now present our results
for pseudo-integrable billiards.

\par Figs.~(8) and (9) show plots of $P(s)$ and $\Delta_3(L)$
for a non-degenerate L-shaped billiard.  A total of 3000 
lengths have been considered after excluding the shortest
3000 orbits. The correlations are clearly Poissonian
as in case of integrable or chaotic billiards. 

\par For the 3-step billiard where fewer lengths  are
available (about 1250), we have carried out a similar study and the
results are shown in Figs.~(10) and (11). Deviations
from the Possion behaviour can now be seen especially in the
spectral rigidity. By considering shorter lengths in the
L-shaped billiard, similar deviations were observed. 

\par The statistical properties of fluctuations in 
the length spectrum of pseudo-integrable systems 
are thus similar to those of 
chaotic billiards. For long orbits the correlations
are Poisson while deviations exist for shorter
orbits.

\section{Discussions and Conclusions}
\label{sec:concl}

\par Quantum billiards are experimentally realizable
in the form of microwave cavities since the wave
equations are identical \cite{Stockman,Sridhar}.
They are also interesting in their own right 
and have proved to be the testing grounds for
several ideas in the field of Quantum Chaos.
Of all possible shapes, polygonal 
billiards have perhaps been the least understood 
largely because very little was known about the
organisation of periodic orbits. The results
presented here have however been used recently
to obtain convergent semiclassical eigenvalues
with arbitrary non-periodic trajectories \cite{prl4?}
as well to demonstrate that periodic orbits
provide excellent estimates of two-point correlations
in the quantum energy spectrum \cite{rapid?}.
Though it is beyond the scope of this article to
review these recent developments on quantum polygonal 
billiards, we refer the interested 
reader to these articles and the references contained
therein.

\par In the previous sections we have explored the organisation
of periodic orbits in rational polygonal billiards that
are pseudo-integrable. Our main conclusions are as follows :

\bigskip
 
\par $\bullet$ Orbit families obey the sum rule $\langle \sum_p \sum_{r=1}
^{\infty} a_p\delta(l-rl_p)/(rl_p)\rangle = 2\pi b_0$ thereby
giving rise to the proliferation law $N(l) = \pi b_0 l^2/ \langle
a(l) \rangle$ for all rational polygons.

\par $\bullet$ The quantity $b_0$ is approximately $1/N$ in
generic generic rational billiards and deviations from 
this value are observed to be significant in degenerate
situations.  

\par $\bullet$ $\langle a(l) \rangle$ increases initially
before saturating to a value much smaller than the maximum
allowed area $NA$.
The asymptotic proliferation law is thus
quadratic even for systems that are not almost-integrable and
the density of periodic orbits lengths is far greater than
an equivalent integrable systems having the same area.

\par $\bullet$ The normalised average area $\langle a(l) \rangle /NA$
decreases with the genus of the billiard while $b_0$ is
approximately 0.25. Periodic orbits thus proliferate faster
with an increase in genus. 

\par $\bullet$ The statistical properties of the fluctuations in the
length spectrum are Poisson when
the orbits considered are long.

\section{Acknowledgements}

\par Part of the work reviewed here was published earlier in
collaboration with Sudeshna Sinha and it a pleasure to
acknowledge the numerous discussions we had on this subject.
I have also benifitted directly or indirectly from 
interesting discussions
with Predrag Cvitanovi\'{c}, Gregor Tanner and Bertrand Gerogeot.


\begin{references}


\bibitem{email} email : biswas@kaos.nbi.dk
\bibitem{permanent} On leave from the Theoretical Physics Division,
Bhabha Atomic Research Centre, Bombay 400 085, India.
\bibitem{PJ} P.J.Richens and M.V.Berry, Physica D{\bf 2}, 495(1981).
\bibitem{rem0} The few examples of integrable polygons are the
rectangle and triangles with internal angles ($\pi/3,\pi/3,\pi/3)$),
($\pi/3,\pi/6,\pi/2$) and ($\pi/4,\pi/4,\pi/2$).
\bibitem{MC} M.C.Gutzwiller, {\it Chaos in Classical and Quantum Mechanics},
Springer, New York, 1990.
\bibitem{gutkin} E.Gutkin, Physica D{\bf 19},311(1986); J. Stat. Phys
{\bf 83}, 7 (1996).
\bibitem{masur} H.Masur, Ergod. Th. and Dyn. Sys. {\bf 10},
151(1990); MSRI,215(1988).
\bibitem{veech} W.A.Veech, Geo. and Func. Anal. {\bf 2}, 341(1992).
\bibitem{vega} J.Vega, T.Uzer and J.Ford, Phys. Rev. E{\bf 48}, 3414 (1993).
\bibitem{cc} T.Cheon and T.D.Cohen, Phys. Rev. Lett. 
{\bf 62},2769(19891).
\bibitem{PLA} D.Biswas and S.Sinha, Phys. Lett. A, {\bf 173}, 392 (1993). 
\bibitem{PCBE} P.Cvitanovi\'{c} and B.Eckhardt,
J. Phys. A. {\bf 24}, L237 (1991).
\bibitem{HO} J.H.Hannay and A.M.Ozorio de Almeida,
J.Phys. A{\bf 17},3429(1984).
\bibitem{keller} J.B.Keller and S.I.Rubinow, Ann. Phys. (N.Y.)
{\bf 9}, 24(1960).
\bibitem{neglect-isolated1} The respective contributions
of isolated and diffractive periodic orbits are
$O(k^{-1})$ and $O(k^{-1-\nu/2})$ where $\nu$ is the total
number of (singular) vertex connections in a diffractive
periodic orbit \cite{niall}.
\bibitem{niall} N.D.Whelan, Phys. Rev. E {\bf 51},3778(1995);
N. Pavloff and C. Schmit, Phys. Rev. Lett. {\bf 75},61(1995).
\bibitem{prl1} D.Biswas and S.Sinha, Phys. Rev. Lett. {\bf 70}, 916(1993)
\bibitem{rapid?} D.Biswas, Phys. Rev. E, {\bf 54}, R1044(1996). 
\bibitem{pramana} D.Biswas, Pramana - Journal of Physics, {\bf 42},
447 (1994).
\bibitem{BV} A.Selberg, J. Indian Math. Soc. {\bf 20}, 47 (1956); 
N.Balazs and A.Voros, Phys. Rep. {\bf 143}, 109 (1986). 
\bibitem{berry85} M.V.Berry, Proc. Roy. Soc. Lond. Ser. A{\bf 400},
229(1985). 
\bibitem{prl2} D.Biswas, Phys. Rev. Lett. {\bf 71}, 2714 (1993).
\bibitem{Stockman} H.-J. Stockmann and J. Stein, Phys. Rev. Lett. {\bf 64},
2867 (1990).
\bibitem{Sridhar} S.Sridhar, Phys. Rev. Lett. {\bf 67}, 785 (1991). 
\bibitem{prl4?} D.Biswas, {\it Quasi-Classical Quantization}, to
be published. 

\end{references}
\end{document}